\documentstyle[12pt]{article}
\newcommand{\Mb}{\bar{M}}
\newcommand{\Nb}{\bar{N}}
\newcommand{\Dirac}{ D_A}

\newcommand{\ba}{\begin{eqnarray}}
\newcommand{\na}{\end{eqnarray}}
\newcommand{\ban}{\begin{eqnarray*}}
\newcommand{\nan}{\end{eqnarray*}}

\newcommand{\ho}{\widehat O}

\begin{document}
\title{Topological Quantum Field Theory and \\
Seiberg-Witten Monopoles}
\author{\small R. B. Zhang , B. L. Wang, A. L. Carey  \\
\small Department of Pure Mathematics  \\
\small J. McCarthy \\
\small Department of Physics and Mathematical physics\\
\small University of Adelaide\\
\small Adelaide, S. A.,  Australia}
\date{ }
\maketitle

\begin{abstract}
A topological quantum field theory is introduced which
reproduces the Seiberg-Witten invariants of four-manifolds.
Dimensional reduction of this topological field theory leads
to a new one in three dimensions. Its partition function
yields a three-manifold invariant, which can be regarded as
the Seiberg-Witten version of Casson's invariant. A Geometrical
interpretation of the three  dimensional quantum field theory is
also given.
\end{abstract}

\vspace{3cm}
\noindent
\section{\small INTRODUCTION}
The field of low dimensional geometry and topology\cite{Atiyah}
has undergone a dramatic phase of progress in recent years, prompted,
to a large extend,  by new ideas and discoveries in
mathematical physics. The discovery of quantum groups\cite{D-J} in
the study of the Yang-Baxter equation\cite{Baxter} has reshaped the
theory of knots and links\cite{Jones}\cite{RT}\cite{ZGB};
the study of conformal
field theory and quantum Chern-Simons theory\cite{W3}
in physics had a profound  impact on the theory of three-manifolds;
and most importantly, investigations
of the classical Yang-Mills theory led to the creation
of the Donaldson theory of four-manifolds\cite{F-U}\cite{Donaldson}.
Very recently, Witten\cite{W1} discovered a new set of invariants
of four-manifolds in the study of the Seiberg-Witten
monopole equations, which have their origin in supersymmetric
gauge theory. The Seiberg-Witten theory, while closely related
to Donaldson theory, is much easier to handle.
Using Seiberg-Witten theory,
proofs of many theorems in Donaldson theory have been simplified,
and several important new results have also been obtained\cite{Tau1}
\cite{Tau2}\cite{K-M}.
However, being only a several months old,
Seiberg-Witten theory remains to be fully developed.
Solutions to any of the outstanding problems, e.g., the precise
relationship to Donaldson theory, will undoubtedly
have great impact on the theory of four-manifolds.

As is well known, Donaldson theory can be cast  into the
framework of topological quantum field theory\cite{W2}\cite{Thompson}.
This formulation
provides a useful tool for exploring conceptual aspects
of the theory, and in some cases, e.g., for Kahler manifolds,
it even enables explicit computations of the Donaldson invariants.

Our aim here is to study the Seiberg-Witten invariants and
related three-manifold invariants using topological quantum
field theory techniques.  We first construct a four dimensional
topological quantum field theory which reproduces the Seiberg-
Witten invariants as correlation functions, and also examine
some of its properties.  Then we dimensionally
reduce it to  obtain a three-dimensional  topological quantum
field theory, the correlation functions of which give rise to
invariants of three-manifolds. The partition function
of the three-dimensional theory is study in some detail;
in particular, we cast it into the Atiyah-Jeffrey framework,
thus to provide it with an interpretation in pure geometrical
terms.\footnote{While this paper was being finished we
noticed that Labastida and Mari\~no \cite{Labast}
 also obtained the siminar result in
4-dimensional case.}

\section{\small TOPOLOGICAL FIELD THEORY}
\subsection{Seiberg-Witten invariants and monopole equations}
The Seiberg-Witten monopole equations are classical field theoretical
equations involving a $U(1)$ gauge field and a complex Weyl spinor on a
four dimensional manifold.   Let $X$ denote the four-manifold, which
is assumed to be oriented and closed.  If $X$ is spin,  there exist
positive and negative spin bundles $S^\pm$ of rank two. Introduce a
complex line bundle $L\rightarrow X$. Let $A$ be a connection on
$L$ and $M$ be a section of the product bundle $S^+\otimes L$.
The Seiberg-Witten monopole equations read
\ba
F_{k l}^+&=&-{{i}\over{2}}\Mb\Gamma_{k l}M,\nonumber\\
\Dirac M&=&0,   \label{master}
\na
where $\Dirac$ is the twisted Dirac operator, $\Gamma_{ i j}
={{1}\over{2}}[\gamma_i, \gamma_j]$, and $F^+$ represents the
self-dual part of the curvature of $L$ with connection $A$.

If $X$ is not a spin manifold, then spin bundles do not exist.
However, it is always possible to introduce the so called
$Spin_c$ bundles $S^\pm\otimes L$, with $L^2$ being a line bundle.
Then in this more general setting, the Seiberg-Witten
monopoles equations look formally the same as (\ref{master}),
but the $M$ should  be interpreted as
a section of the the $Spin_C$ bundle $S^+\otimes L$.

Denote by $\cal M$ the moduli space of solutions of the
Seiberg-Witten monopole equations up to gauge transformations.
Generically, this space is a manifold. Its virtual
dimension is equal to the number of solutions of the following
equations
\ba
(d\psi)^+_{k l} + {{i}\over{2}}
\left(\Mb\Gamma_{k l}N+ \Nb\Gamma_{k l}M\right)&=&0,\nonumber\\
\Dirac N + \psi M&=&0, \nonumber \\
\nabla_k\psi^k + {{i}\over{2}}(\overline N M - \overline M N)
&=&0,   \label{deform}
\na
where $A$ and $M$ are a given solution of (\ref{master}),
$\psi\in \Omega^1(X)$ is a one form, $(d\psi)^+
\in \Omega^{2, +}(X)$ is the self dual part of the two
form $d\psi$, and $N\in S^+\otimes L$.
The first two of the equations in (\ref{deform}) are
the linearization of the monopole equations (\ref{master}),
while the last one is a gauge fixing condition.
Though with a rather unusual form, it arises naturally
from the dual operator governing gauge transformations
\ban
C: \Omega^0(X)&\rightarrow&\Omega^1(X)\oplus (S^+\otimes L)\\
\phi  &\mapsto& (- d \phi, i\phi M).
\nan
Let
\ba
T:  \Omega^1(X)\oplus(S^+\otimes L)\rightarrow
 \Omega^0(X)\oplus \Omega^{2, +}(X)\oplus(S^-\otimes L),
\na
be the operator governing  equation (\ref{deform}),
namely, the operator which allows us to rewrite (\ref{deform}) as
\ban
T(\psi, N)&=&0.
\nan
Then $T$ is an elliptic operator, the index
$Ind(T)$ of which yields the virtual dimension of $\cal M$.
A straightforward application of the Atiyah-Singer index
theorem gives
\ban
Ind(T) &=& -{ {2\chi(X)+3\sigma(X)}\over{4}} + c_1(L)^2,
\nan
where
$\chi(X)$ is the Euler character of $X$, $\sigma(X)$ its
signature index and $c_1(L)^2$ is the square of the
first Chern class of $L$ evaluated on $X$ in the standard way.

When $Ind(T)$ equals zero, the moduli space generically
consists of a finite number of points, $\cal M=\{p_t: t=1, 2, ..., I\}$.
Let $\epsilon_t$ denote the sign of the determinant of the
operator $T$ at $p_t$, which can be defined with mathematical rigour.
Then the Seiberg-Witten invariant of
the four-manifold $X$ is defined by
\ba
\sum_1^I\epsilon_t.
\na

The fact that this is indeed an invariant(i.e., independent of
the metric) of $X$ is not very difficult to prove, and we refer to
\cite{W1} for details.  As a matter of fact,  the number of solutions of a
system of equations  weighted by the sign of the operator
governing the equations(i.e., the analog of $T$) is
a topological invariant in general\cite{W1}. This point of view
has been extensively explored by Vafa and Witten\cite{V-W}
within the framework of topological quantum field theory in
connection with the so called $S$ duality.  Here we wish to
explore the Seiberg-Witten invariants following a similar
line as that taken in \cite{W2}\cite{V-W}.

\subsection{Topological Lagrangian}
Introduce a Lie superalgebra with an odd generator $Q$
and two even generators  $U$ and $\delta$ obeying the following
(anti)commutation relations
\ba
[U, Q]=Q, & [Q, Q]=2\delta,&  [Q, \delta]=0.\label{superalgebra}
\na
We will call $U$ the ghost number operator, and $Q$ the BRST
opearator.

Let $A$ be a connection of $L$ and $M\in S^+\otimes L$. We define
the action of the superalgebra on these fields by requiring that
$\delta$ coincide with a gauge transformation with a gauge parameter
$\phi\in\Omega^0(X)$.  The field multiplets associated
with $A$ and $M$ furnishing representations of the superalgebra
are $(A, \psi, \phi)$, and $(M, N)$, where
$\psi\in\Omega^1(X)$, $\phi\in\Omega^0(X)$, and $N$ is a section
of $S^+\otimes L$. They transform under the action of
the superalgebra according to
\ba
[Q, A_i]=\psi_i,              & [Q, M]=N,\nonumber \\
{[}Q, \psi_i]=-\partial_i \phi, &  [Q, N]=i\phi M, \nonumber\\
{[}Q, \phi]=0.
\na
We assume that both $A$ and $M$ have ghost number $0$, and thus will
be regarded as bosonic fields when we study their quantum field theory.
The ghost numbers of other fields can be read off
the above transformation rules. We have that $\psi$ and $N$ are of
ghost number $1$, thus are fermionic, and $\phi$ is of ghost number
$2$ and bosonic. Note that the multiplet $(A, \psi, \phi)$ is what
one would get in the topological field theory for Donaldson invariants
except that our gauge group is $U(1)$, while the existence of $M$ and
$N$ is a new feature. Also note that both $M$ and $\psi$ have the
wrong statistics.

In order to construct a quantum field theory which will reproduce
the Seiberg-Witten invariants as correlation functions,
anti-ghosts and Lagrangian multipliers are also required.
We introduce the anti-ghost multiplet $(\lambda, \eta)$ $\in \Omega^0(X)$,
 such that
\ba
[U, \lambda]=-2\lambda, &  [Q, \lambda]=\eta,& [Q, \eta]=0,
\na
and the Lagrangian multipliers $(\chi, H)\in \Omega^{2, +}(X)$,
and $(\mu, \nu)\in S^-\otimes L$ such that
\ba
[U, \chi]=-\chi, & [Q, \chi]=H, & [Q, H]=0;  \nonumber \\
{[}U, \mu]=-\mu, & [Q, \mu]=\nu, & [Q, \nu]=i\phi \mu.
\na

With the given fields, we construct the following functional which
has ghost number $-1$:
\ba
V&=&\int_X \left\{[\nabla_k\psi^k + {{i}\over{2}}
(\overline N M - \overline M N)] \lambda -
\chi^{k l}\left(H_{k l} -F_{k l}^+
 -{{i}\over{2}}\Mb\Gamma_{k l}M\right)\right.\nonumber\\
&-&\left.{\bar\mu}\left(\nu-i\Dirac M\right)
-{\overline{\left(\nu - i\Dirac M\right)}}\mu
\right\},\label{V-S}
\na
where the indices of the tensorial fields are raised and lowered by
a given metric $g$ on $X$, and the integration measure is the standard
$\sqrt{g} d^4x$. Also, $\overline M$ and $\bar \mu$ etc.
represent the hermitian conjugate of the spinorial fields.
In a formal language,  ${\overline M}\in S^+\otimes L^{-1}$ and
$\bar\mu, \bar\nu, \overline{\Dirac M}$ $\in S^-\otimes L^{-1}$.
Following the standard procedure in constructing topological quantum
field theory, we take the classical action of our theory to be
\ba
S&=&[Q, V],  \label{action}
\na
which has ghost number $0$.   One can
easily show that $S$ is also  BRST invariant,
i.e., \ba [Q, S]&=&0, \na
thus it is invariant under the full superalgebra(\ref{superalgebra}).

The bosonic Lagrangian multiplier fields $H$ and $\nu$ do not have
any dynamics, and so  can be eliminated from the action by using their
equations of motion
\ba
H_{k l} &=&{{1}\over{2}}\left(F_{k l}^++{{i}\over{2}}\Mb\Gamma_{k l}M\right),
\nonumber \\
\nu&=&{{1}\over{2}}i\Dirac M.  \label{motion}
\na
Then we arrive at the following expression for the action
\ba
S&=&\int_X\left\{[-\Delta\phi +\overline M M\phi - i \overline N N] \lambda
- [\nabla_k\psi^k + {{i}\over{2}}(\overline N M - \overline M N)]\eta
+2 i \phi{\bar\mu} \mu \right.\nonumber \\
&+&\overline{\left(i\Dirac N - \gamma.\psi M\right)}\mu
-\bar{\mu}\left(i\Dirac N - \gamma.\psi M\right)\nonumber\\
&-&\left.\chi^{k l}\left[\left(\nabla_k\psi^l-\nabla_l\psi^k\right)^+
+{{i}\over{2}}\left(\Mb\Gamma_{k l}N+\Nb\Gamma_{k l}M\right)\right]\right\}
\nonumber\\
&+& S_0,  \label{S}
\na
where $S_0$ is given by
\ba
S_0&=& \int_X\left\{ {{1}\over{4}}|F^+ + {{i}\over{2}}\Mb\Gamma M|^2
+{{1}\over{2}}|\Dirac M|^2\right\}.
\na
It is interesting to observe that $S_0$ is nonnegative, and vanishes
if and only if $A$ and $M$ satisfy the Seiberg-Witten monopole
equations.  As pointed out in \cite{W1}, $S_0$ can be rewritten as
\ban
S_0&=& \int_X\left\{ {{1}\over{4}}|F^+|^2 + {{1}\over{4}} |M|^4
+{{1}\over{8}}R|M|^2 + g^{i j} \overline{D_iM}D_j M\right\},
\nan
where $R$ is the scalar curvature of $X$ associated with the metric $g$.
If $R$ is nonnegative over the entire $X$, then the only square integrable
solution of the monopole equations (\ref{master}) is $A$ is a anti-self-dual
connection  and $M=0$.

\subsection{Quantum theory}
We will now investigate the quantum field theory defined by the classical
action (\ref{S}) with the path integral method.
Let $\cal F$ collectively denote all the fields. The partition function
of the theory is defined by
\ban
Z&=&\int \cal{D F} \exp(-{{1}\over{e^2}}S),
\nan
where $e\in\bf R$ is the coupling constant.
The integration measure $\cal{D F}$ is
defined on the space of all the fields. However, since $S$ is
invariant under the gauge transformations, we assume the integration
over the gauge field to be performed over the gauge orbits of $A$.
In other words, we fix a gauge for the $A$ field using, say,
a Faddeev-Popov type procedure.  This can be carried out in
the standard manner, thus there is no need for us to spell
out the details here.
The integration measure $\cal{D F}$ can be shown to be invariant
under the super charge $Q$. Also, it does not explicitly
involve the metric $g$ of $X$.

Let $W$ be any operator in the theory. Its correlation function
is defined by
\ban
Z[W]&=&\int \cal{D F} \exp(-{{1}\over{e^2}}S) W.
\nan
It follows from the $Q$ invariance of both the action $S$ and the path
integration measure that for any operator $W$,
\ban
Z[[Q, W]]&=&\int \cal{D F} \exp(-{{1}\over{e^2}}S)[Q,  W]\\
&=&0.
\nan

For the purpose of constructing topological invariants of the four-
manifold $X$, we are particularly interested in operators $W$
which are BRST closed,
\ba [Q, W]&=&0,\label{in1}\na
but not BRST exact, i.e., can not be expressed as the (anti)-
commutators of $Q$ with other operators.
For such a $W$, if its variation with respect to the metric $g$
is $BRST$ exact,
\ba \delta_g W&=&[Q, W'],\label{in2} \na
then its correlation function $Z[W]$ is a topological invariant of $X$
(by that  we really mean that it does not depend on the metric $g$):
\ban
\delta_gZ[W]&=&\int\cal{D F} \exp(-{{1}\over{e^2}}S)
[Q, W' - {{1}\over{e^2}}\delta_gV.  W]  \\
&=&0.
\nan
In particular, the partition function $Z$ itself is a topological invariant.
In fact, under certain conditions imposed on $X$, the partition function
coincides with the Seiberg-Witten invariants, as we will prove below.

Another important property of the partition function is that it does
not depend on the coupling constant $e$:
\ban
{{\partial Z}\over{\partial e^2} } &=&\int\cal{D F}\
{{1}\over{e^4}} \exp(-{{1}\over{e^2}}S)[Q, V]  \\
&=&0.
\nan
Therefore, $Z$ can be computed {\em exactly} in the limit when the coupling
constant goes to zero.  Such a computation can be carried out in the
standard way:  Let $A^o$, $M^o$ be a solution of the equations of motion
of $A$ and $M$ arising from the action $S$.  We expand the fields $A$ and
$M$ around this classical configuration,
\ban
A=A^o + e a,&  M=M^o + e m,
\nan
where $a$ and $m$ are the quantum fluctuations of $A$ and $M$ respectively.
All the other fields do not acquire background components,
thus are purely quantum mechanical. We scale them by the coupling
constant $e$, by  setting  $N$ to $e N$, $\phi$ to $e\phi$ etc..
To the order $o(1)$ in $e^2$, we have
\ba
Z&=&\sum_{p}\exp(-{{1}\over{e^2}}S^{(p)}_{cl})\int\cal{D F'} \exp(-S^{(p)}_q),
\label{Z}
\na
where $S_q^{(p)}$ is the quadratic part of the action in the quantum
fields and depends on the gauge orbit of the
classical configuration $A^o$, $M^o$, which we label
by $p$.  Explicitly,
\ban
S_q^{(p)}&=&\int_X\left\{[-\Delta\phi +\overline M^o M^o\phi -
i\overline N N]\lambda
- [\nabla_k\psi^k + {{i}\over{2}}(\overline N M^o - \overline M^o N)]\eta
+2 i \phi{\bar\mu} \mu \right.\nonumber \\
&+&\overline{\left(i D_{A^o} N - \gamma.\psi M^o\right)}\mu
-\bar{\mu}\left(i D_{A^o} N - \gamma.\psi M^o\right)\nonumber\\
&-&\chi^{k l}\left[\left(\nabla_k\psi^l-\nabla_l\psi^k\right)^+
+{{i}\over{2}}\left(\Mb^o\Gamma_{k l}N+\Nb\Gamma_{k l}M^o\right)\right]
\nonumber\\
&+&\left.{{1}\over{4}}|f^++{{i}\over{2}}(\bar{m}\Gamma M^o+\Mb^o\Gamma m)|^2
+{{1}\over{2}}|i D_{A^o}m + \gamma.a M^o|^2\right\},
\nan
with $f^+$ the self -dual part of $f=d a$.
The classical part of the action is given by
\ban
S_{cl}^{(p)}&=&S_0|_{A=A^o, M=M^o}.
\nan
The integration measure $\cal{D F'}$  has exactly the same form as
$\cal{D F}$ but with $A$ replaced by $a$, and $M$ by $m$,
$\bar M$ by $\bar m$ respectively.
Needless to say, the summation over $p$ runs through  all gauge
classes of  classical configurations.

Let us now examine further features of our quantum field theory.
A gauge class of classical configurations may give a non-zero
contribution to the partition function in the limit
$e^2\rightarrow 0$ only
if $S_{cl}^{(p)}$ vanishes, and this happens if and only if
$A^o$ and $M^o$ satisfy (\ref{master}).  Therefore,
the Seiberg-Witten monopole equations are recovered from
the quantum field theory.

The equations of motion of the fields $\psi$ and $N$
in the semi-classical approximation can be easily derived from the
quadratic action $S^{(p)}_q$, solutions of which are the zero modes
of the quantum fields $\psi$ and $N$.  The equations  of motion read
\ba
(d\psi)^+_{k l} + {{i}\over{2}}
\left(\Mb^o\Gamma_{k l}N+ \Nb\Gamma_{k l}M^o\right)&=&0,\nonumber\\
D_{A^o} N + {\gamma.\psi}M^0&=&0, \nonumber \\
\nabla_k\psi^k + {{i}\over{2}}(\overline N M- \overline M N)&=&0.
\label{moduli}
\na
Note that they are exactly the same equations which we have already
discussed in (\ref{deform}).  The first two equations are
the linearization of the monopole equations, while the last is a
`gauge fixing condition' for $\psi$.  The dimension of the space of
solutions of these equations is the virtual dimension of the moduli space
$\cal M$. Thus, within the context of our quantum field theoretical
model, the virtual dimension of $\cal M$ is identified with the
number of the zero modes of the quantum fields $\psi$ and $N$.

For simplicity we assume that there are no zero modes of $\psi$ and $N$,
i.e., the moduli space is zero dimensional. Then no zero modes exist for
the other two fermionic fields $\chi$ and $\mu$.
To compute the partition function in this case, we first observe that
the quadratic action $S^{(p)}_q$ is invariant under the supersymmetry
obtained by expanding $Q$ to first order in the quantum fields
around the monopole solution $A^o$, $M^o$ (equations of motion for
the nonpropagating fields $H$ and $\nu$ should also be used.). This
supersymmetry transforms the set of $8$ real bosonic fields (each complex
field is counted as two real ones; the $a_i$ contribute $2$ upon gauge
fixing.) and  the set of $16$ fermionic fields to each other. Thus at a
given monopole background we obtain
\ban
\int\cal{D F'} \exp(-S^{(p)}_q) &=& {{ {\rm Pfaff}(\nabla_F)}\over
{|{\rm Pfaff} (\nabla_F)|}}=\epsilon^{(p)},
\nan
where $\epsilon^{(p)}$ is $+1$ or $-1$.
In the above equation,
$\nabla_F$ is the skew symmetric first order differential
operator defining the fermionic part of the action $S^{(p)}_q$,
which can be read off from $S^{(p)}_q$ to be
$\nabla_F=\left(\begin{array}{lr}
0& T\\ -T^* &0\end{array}\right)$.
Therefore, $\epsilon^{(p)}$  is the sign of
the determinant of the elliptic operator $T$ at the monopole background
$A^o$, $M^o$, and the partition function
\ba
Z&=&\sum_{p}\epsilon^{(p)},
\na
coincides with the Seiberg-Witten invariant of the four-manifold $X$.

When the dimension of the moduli space $\cal M$ is greater than zero,
the partition function $Z$ vanishes identically, due to integration
over zero modes of the fermionic fields.  In order to obtain any non trivial
topological invariants for the  underlying manifold $X$, we need
to examine correlations functions of operators satisfying equations
(\ref{in1}) and (\ref{in2}).  A class of such operators can be
constructed following the standard procedure\cite{W1}.  We define the
following set of operators
\ba
W_{k, 0}&=& {{\phi^k}\over{k!}}, \nonumber\\
W_{k, 1}&=& \psi W_{k-1, 0}, \nonumber\\
W_{k, 2}&=& F W_{k-1, 0}
         - {{1}\over{2}}\psi\wedge\psi W_{k-2, 0}, \nonumber\\
W_{k, 3}&=& F\wedge\psi W_{k-2, 0}
         - {{1}\over{3!}}\psi\wedge\psi\wedge\psi W_{k-3, 0}, \nonumber\\
W_{k, 4}&=& {{1}\over{2}} F\wedge F W_{k-2, 0}
         - {{1}\over{2}}F\wedge\psi\wedge\psi W_{k-3, 0}
         -{{1}\over{4!}} \psi\wedge\psi\wedge\psi\wedge\psi W_{k-4, 0}.
\na
These operators are clearly independent of the metric $g$ of $X$.
Although they are not BRST invariant except for $W_{k, 0}$,
they obey the following equations
\ban
d W_{k, 0}&=&-[Q, W_{k, 1}], \\
d W_{k, 1}&=& [Q, W_{k, 2}], \\
d W_{k, 2}&=&-[Q, W_{k, 3}],\\
d W_{k, 3}&=& [Q, W_{k, 4}],\\
d W_{k, 4}&=&0,
\nan
which allow us to construct BRST invariant operators from the the $W$'s
in the following way: Let $X_i$,  $i=1, 2, 3$,  $X_4=X$,  be compact
manifolds without boundary embedded in $X$. We assume
that these submanifolds are homologically nontrivial. Define
\ba
\ho_{k, 0}&=&W_{k, 0},\nonumber\\
\ho_{k, i}&=&\int_{X_i} W_{k, i},  \ \ \ \ \ \  i=1, 2, 3, 4.
\na
As we have already  pointed out, $\ho_{k, 0}$ is BRST invariant.
It follows from the descendent equations that
\ban
[Q, \ho_{k, i}] &=&\int_{X_i}[Q,  W_{k, i}]\\
&=&\int_{X_i}d W_{k, i-1}=0.
\nan
Therefore the operators $\ho$ indeed have the properties (\ref{in1})
and (\ref{in2}).  Also, for the boundary $\partial K$ of an $i+1$
dimensional manifold $K$ embedded in $X$,  we have
\ban
\int_{\partial K}  W_{k, i}&=& \int_{K} d  W_{k, i}\\
          &=&[Q, \int_{K}   W_{k, i+1}],
\nan
is BRST trivial. The correlation function of $\int_{\partial K}  W_{k, i}$
with any BRST invariant operator is identically zero.
This in particular shows that the $\ho$'s only depend on the homological
classes of the submanifolds $X_i$.

\section{\small DIMENSIONAL REDUCTION}
In this section we dimensionally reduce the quantum field
theoretical model for the Seiberg-Witten invariant from
four dimensions to three dimensions, thus to obtain a new
topological quantum field theory defined on $3$- manifolds.
Its  partition function yields a $3$- manifold invariant,
which can be regarded as the Seiberg-Witten version of
Casson's invariant  \cite{Casson}\cite{Tau3}.

\subsection{Three-dimensional field theory}
We take the four-manifold $X$ to be of the form
$Y\times [0,\ 1]$ with $Y$ being a compact
$3$-manifold without boundary.  The metric on $X$ will be taken to
be $$(d s)^2 = (d t)^2 + \sum_{ i, j}^3 g_{i j}(x) d x^i d x^j,$$
where the `time' $t$-independent $g(x)$ is the Riemannian metric on $Y$.
We assume that $Y$ admits a spin structure which is compatible with the
$Spin_c$ structure of $X$, i.e., if we think of $Y$ as embedded in $X$, then
this embedding induces maps from the $Spin_c$ bundles $S^\pm \otimes L$
of $X$ to $\tilde S\otimes L$, where $\tilde S$ is a spin bundle
and $L$ is a line bundle over $Y$.

To perform the dimensional reduction, we impose the condition that all
fields are $t$ in dependent. This leads to the following action
\ba
S&=&\int\sqrt{g}d^3x \left\{ [-\Delta\phi +\overline M M\phi
- i \overline N N] \lambda
- [\nabla_k\psi^k + {{i}\over{2}}(\overline N M - \overline M N)]\eta
+2 i \phi{\bar\mu} \mu \right.\nonumber \\
&+&\overline{\left[ i(\Dirac +b) N - (\sigma.\psi -\tau)M\right]}\mu
-\bar{\mu}\left[ i(\Dirac +b) N - (\sigma.\psi -\tau)M\right]    \nonumber\\
&-&2\chi^k\left[-\partial_k\tau + *(\nabla\psi)_k
-\Mb\sigma_k N-\Nb\sigma_k M\right] \nonumber\\
&+&\left. {{1}\over{4}}|*F -\partial b -\Mb\sigma M|^2
+{{1}\over{2}}|(\Dirac+b)M|^2\right\}, \label{S3}
\na
where the $k$ is a three-dimensional index, and $\sigma_k$ are the
Pauli matrices. The fields $b, \tau\in\Omega^0(Y)$ respectively
arose from $A_0$ and $\psi_0$ of the four dimensional theory,
while the meanings of the other fields are clear.  The BRST symmetry
in four-dimensions carries over to the three-dimensional theory.
The BRST transformations rules for $(A_i, \psi_i, \phi)$, $i=1, 2, 3$,
$(M, N)$, and $(\lambda, \eta)$ are the same as before, but
for the other fields, we have
\ba
[Q, b]&=&\tau,   \nonumber\\
{[}Q, \tau]&=&0,\nonumber\\
{[Q}, \chi_k]&=&{{1}\over{2}}\left(*F_k -\partial_k b
              -\Mb\sigma_k M\right), \nonumber\\
{[Q}, \mu]&=&{{1}\over{2}} i (\Dirac+b)M.
\na

The action $S$ is cohomological in the sense that $S=[Q, V_3]$,
with $V_3$ being the dimensionally reduced version of $V$ defined
by (\ref{V-S}), and $[Q, S]=0$. Thus it gives rise to a topological
field theory upon quantization.  The partition function of the
theory
\ban
Z&=&\int \cal{D F} \exp(-{{1}\over{e^2}}S),
\nan
can be computed exactly in the limit $e^2\rightarrow 0$, as it is coupling
constant independent. We have, as before,
\ban
Z&=&\sum_{p}\exp(-{{1}\over{e^2}}S^{(p)}_{cl})\int\cal{D F'} \exp(-S^{(p)}_q),
\nan
where $S^{(p)}_q$ is the quadratic part of $S$ expanded around a classical
configuration with the classical parts for the fields $A, M, b$
being $A^o, M^o, b^o$, while those for all the other fields being
zero. The classical action $S^{(p)}_{cl}$ is given by
\ban
S^{(p)}_{cl}&=&\int_Y\left\{{{1}\over{4}}|*F^o- d  b^o-\Mb^o\sigma M^o|^2
+{{1}\over{2}}|(D_{A^o}+b^o)M^o|^2\right\},
\nan
which can be rewritten as
\ban
S^{(p)}_{cl}&=&\int_Y\left\{ {{1}\over{4}}|*F^o -\Mb^o\sigma M^o|^2
+{{1}\over{2}}|D_{A^o} M^o|^2
+{{1}\over{2}}| d  b^o|^2 +{{1}\over{2}} |b^o M^o|^2\right\}.
\nan

In order for the classical configuration to have nonvanishing
contributions to the partition function, all the terms in $S^{(p)}_{cl}$
should vanish seperately. Therefore,
\ba
*F^o-\Mb^o\sigma M^o&=&0, \nonumber\\
D_{A^o}M^o&=&0, \label{dim3} \na
and
\ban
b^o&=&0,
\nan
where the last condition requires some explanation.
When we have a trivial solution of the equations (\ref{dim3}),
it can be replaced by the less stringent condition $ d  b^o=0$.
However, in a more rigorous treatment of the problem at hand, we
in general perturb the equations (\ref{dim3}), then
the trivial solution does not arise.

Let us define an operator
\ba
\tilde{T}:   \Omega^0(Y)\oplus \Omega^1(Y)\oplus(\tilde S\otimes L)&\rightarrow
&\Omega^0(Y)\oplus \Omega^1(Y)\oplus(\tilde S\otimes L),\nonumber\\
(\tau, \psi, N)\mapsto(-d^*\psi+{{i}\over{2}}(\overline N M - \overline M N),
&& *(d\psi) - d\tau-\bar N\sigma M -\overline{M}\sigma N,\nonumber  \\
&&  iD_{A}N - (\sigma.\psi -\tau)M),
\na
where the complex bundle $\tilde S\otimes L$ should be regarded as a real
one with twice the rank. This operator is self-adjoint, and is also
obviously elliptic. We will assume that it is Fredholm as well.
In terms of $\tilde{T}$, the equations of motion of
the fields $\chi^i$ and $\mu$ can be expressed as
\ba
\tilde{T}^{(p)}(\tau, \psi, N)&=&0,
\na
where $\tilde{T}^{(p)}$ is the opeartor $\tilde{T}$ with the background
fields $(A^o, M^o)$ belonging  to the gauge class $p$ of classical
configurations .

When the kernel of $\tilde{T}$ is zero, the partition function
$Z$ does not vanish identically. An easy computation leads to
\ban
Z&=&\sum_{p}\epsilon^{(p)},
\nan
where the sum is over all gauge inequivalent solutions of (\ref{dim3}),
and $\epsilon^{(p)}$ is the sign of the determinant of $\tilde T^{(p)}$.

A rigorous definition of the sign of the $det(\tilde T)$ can be
devised. However, if we are to compute only the absolute value of
$Z$, then it is sufficient to know the sign of $det(\tilde T)$
relative to a fixed gauge class of classical configurations.
This can be achieved using the $mod-2$ spectral flow
of a family of Fredholm operators $\tilde{T_t}$ along a path of
solutions of (\ref{dim3}).  More explicitly,
let $(A^o, M^o)$ belong to the gauge class of classical configurations
$p$, and  $(\tilde A^o, \tilde M^o)$ in $\tilde p$. We consider
the solution of the Seiberg-Witten equation on
$X=Y\times [0, 1]$ with $A_0=0$ and also satisfying the following
conditions
\ban
(A, M)|_{t=0}&=&(A^o,  M^o), \\
(A, M)|_{t=1}&=& (\tilde A^o,  \tilde M^o).
\nan
Using this solution in $\tilde T$ results in a family of Fredholm operators,
which has zero kernels at $t=0$ and $1$.  The spectral flow
of $\tilde T_t$, denoted by $q(p, \tilde p)$,
is defined to be the number of eigenvalues which
cross zero with a positive slope minus the number which cross zero
with a negative slope.  This number is a well defined quantity, and
is given by the index of the operator
${{\partial}\over{\partial t}}- \tilde T_t$. In terms of the spectral
flow, we have
\ban
{{det(\tilde T^{(p)})}\over{det(\tilde T^{(\tilde p)})}}
&=&(-1)^{q(p, \tilde p)}.
\nan

Equations (\ref{dim3}) can be derived from the functional
\ban
S_{c-s}&=&{{1}\over{2}}\int_Y A\wedge F
+i\int_Y \sqrt{g} d^3x \overline{M}D_{A}M.
\nan
(It is interesting to observe that this is almost the standard Lagrangian
of a $U(1)$ Chern-Simons theory coupled to spinors,
except that we have taken $M$ to have bosonic statistics.)
$S_{c-s}$ is gauge invariant modulous a constant arising from
the Chern-Simons term upon a gauge transformation. Therefore,
$({{\delta S_{c-s}}\over{\delta A}}, {{\delta S_{c-s}}\over{\delta\bar M}})$
defines a vector field on the quotient space of all $U(1)$ connections
$\cal A$ tensored with the $\tilde S\times L$ sections by the $U(1)$
gauge group $\cal G$,
i.e., $\cal W=(\cal A\times(\tilde S\otimes L))/\cal G$.
Solutions of (\ref{dim3}) are zeros of this vector field, and
$\tilde T^{(p)}$ is the Hessian at the point $p\in\cal W$.  Thus
the partition $Z$ is nothing else but the Euler character of
$\cal W$. This geometrical interpretation will be spelt out more
explicitly in the next subsection by re-interpreting the theory
using the Mathai-Quillen formula\cite{Mathai}.

\subsection{Geometrical interpretation}
To elucidate the geometric meaning of the three-dimensional theory
obtained in the last section, we now cast it into the framework
of Atiyah and Jeffrey\cite{A-J}.  Let us briefly recall the
geometric set up of the Mathai-Quillen formula
as reformulated in reference \cite{A-J}. Let $P$ be a Riemannian
manifold of dimension $2m+dim G$, and $G$ be a compact Lie group
acting on $P$ by isometries.  Then $P\rightarrow P/G$ is a principle
bundle. Let $V$ be a $2m$ dimensional real vector space, which furnishes
a representation  $G\rightarrow SO(2m)$. Form the associated vector
bundle $P\times_G V$. Now the Thom form of $P\times_G V$ can be expressed
\ba
U&=& {{\exp(-x^2)}\over{(2\pi)^{dim G} \pi^m} }
\int \exp\left\{{{i\chi\phi\chi }\over{4}} +i\chi d x -
i\langle \delta\nu, \lambda\rangle\right. \nonumber\\
&-&  \langle\phi, R\lambda\rangle
\left.+\langle\nu , \eta\rangle\right\}
\cal D \eta \cal D\chi\cal D\phi\cal D\lambda, \label{Thomo}
\na
where $x=(x^1, ..., x^{2m})$ is the coordinates of $V$,
$\phi$ and $\lambda$ are
bosonic variables in the Lie algebra $g$ of $G$, and
$\eta$ and  $\chi$ are Grassmannian variables valued in the
Lie algebra and the tangent space of the fiber respectively.
In the above equation, $C$ maps any $\eta\in g$ to the element of the vertical
part of $TP$ generated by $\eta$;
$\nu$ is the $g$ - valued one form on $P$  defined by
$\langle \nu(\alpha), \eta\rangle$ $=$ $\langle \alpha , C(\eta )\rangle$,
for all vector fields $\alpha$; and $R=C^*C$.
Also, $\delta$ is the exterior derivative on $P$.

Now we choose a $G$ invariant map  $s: P\rightarrow
V$, and pull back the Thom form $U$. Then the top form
on $P$ in $s^*U$ is the Euler class.  If $\{\delta p\}$ forms a basis of the
cotangent space of $P$(note that $\nu$ and $\delta s$ are one forms on $P$),
we replace it by a set of Grassmannian variables $\{\psi\}$ in $s^*U$,
then intergrate them away. We arrive at
\ba
& & {{1}\over{(2\pi)^{dim G} \pi^m} }
\int \exp\left\{-|s|^2 + {{i \chi\phi\chi }\over{4}} +i\chi \delta s  -
i\langle \delta\nu, \lambda\rangle\right.   \nonumber \\
&-&   \langle\phi, R\lambda\rangle
\left.+\langle\psi, C\eta\rangle\right\}
\cal D \eta \cal D\chi\cal D\phi\cal D\lambda\cal D\psi, \label{Thom}
\na
the precise relationship of which with the Euler character of
$P\times_G V$ is $$\int_P (\ref{Thom})=Vol(G)\chi(P\times_G).$$

It is rather obvious that the action $S$ defined by (\ref{S}) for the
four-dimensional theory can be interpreted as the exponent in the
integrand of (\ref{Thom}), if we identify $P$ with
$\cal A\times \Gamma(W^+)$, and $V$ with $\Omega^{2, +}(X)\times \Gamma(W^-)$,
and set $s=(F^+ + {{i}\over{2}}\Mb\Gamma M, \Dirac M)$. Here $\cal A$ is the
space of all $U(1)$ connections of $det(W^+)$, and $\Gamma(W^\pm)$ are
the sections  of $S^\pm \otimes L$ respectively.

For the three-dimensional theory, we wish to show that the partition
function yields the Euler number of $\cal W$.
However, the tangent bundle of $\cal W$ cannot be regarded as an associated
bundle with the  princibal bundle, for which for the formulae
(\ref{Thomo}) or (\ref{Thom}) can readily apply,
some further work is required.

Let $P$ be the principal bundle over $P/G$,  $V$, $V'$ be two
orthogonal representions of $G$.
Suppose there is an embedding from $P\times _G V'$ to
$P\times _G V$ via a $G$-map $\gamma (p): V'\rightarrow V$ for $p\in P$.
Denote the resulting quotient bundle as $E$.
In order to derive the Thom class for $E$, one needs
to choose a section of $E$, or equivalently,  a $G$-map
$s: P\rightarrow V$ such that $s(p)\in (Im \gamma (p))^\perp$.
Then the Euler class of $E$ can be expressed as $\pi _*\rho^* U$,
where $U$ is the Thom class of $P\times _G V$,
$\rho $ is a $G$-map: $P\times  V ' \rightarrow  P\times V$ defined by
\ban
\rho (p, \tau )&=&(p, \gamma (p)\tau + s(p)),
\nan
and $\pi_*$ is the integration along the fiber for the projection
$\pi : P\times V' \rightarrow P/G$. Explicitly,
\ba
\pi_*\rho^* (U) &=& \int \exp \left\{ - |\gamma (p)\tau +s(p)|^2
+i\chi \phi \chi + i \chi \delta (\gamma (p) \tau + s(p))\right.\nonumber\\
& &-i \langle\delta \nu , \lambda \rangle - \langle \phi ,R \lambda \rangle +
\langle\nu , C\eta \rangle\left.
\right\} \cal D \chi\cal D \phi \cal D\tau \cal D\eta \cal D\lambda
\label{quoThom}
\na

Consider the exact sequence
\ban
0\longrightarrow (\cal A \times \Gamma (W)) \times _{\cal G} \Omega ^0(Y)
\stackrel {j}{\longrightarrow} (\cal A \times
 \Gamma (W))\times _{\cal G} (\Omega ^1(Y)
\times \Gamma (W))
\nan
where $ j_{(A, M)}: b \mapsto (-d b, b M)$.
(We assume that $M\ne 0$).
Then the tangent bundle of $ \cal A \times_ {\cal G} \Gamma (W)$ can be
Regarded as the quotient bundle $$(\cal A \times \Gamma (W))
\times _{\cal G}(\Omega ^1(Y)\times\Gamma (W))/Im(j).	$$
We define a vector field on  $\cal A \times_{\cal G} \Gamma (W)$ by
$$
s(A, M) = (*F_A -\bar M \sigma M, \Dirac M ),
$$
which lies in $(Im j)^{\perp}$:
\ba
\int_Y (*F_A -\bar M \sigma M)\wedge*(- d b)
+ \int_Y\sqrt g d^3x \langle \Dirac M , b  M \rangle &=&0, \label{relation}
\na
where we have used the short hand notation
$\langle M_1, M_2\rangle =
{{1}\over{2}}(\overline M_1 M_2+\overline M_2 M_1)$.

Formally applying the formula (\ref{quoThom}) to
the present infinite dimensional situation, we obtain
the Euler calss $\pi _* \rho ^*(U)$  for the tangent bundle
$T(\cal A\times _{\cal G} \Gamma (W))$,
where $\rho$ is the $\cal G$-invariant map $\rho$ is defined by
\ban
 \rho :  \quad \Omega ^0(Y) \longrightarrow \Omega^1 (Y) \times \Gamma (W),\\
\rho(b) = (-d b + *F_A -\bar M \sigma M, (\Dirac +b) M),
\nan
$\pi$ is the projection
$(\cal A \times \Gamma (W)) \times _{\cal G} \Omega ^0(Y) \longrightarrow
\cal A\times _{\cal G} \Gamma (W) \label{bundle}$,
and  $\pi_*$ signifies the integration along the fiber. Also
$U$ is the Thom form of the bundle
\ban
(\cal A \times \Gamma (W))\times_{\cal G} (\Omega ^1(Y)
\times \Gamma (W) ) \longrightarrow \cal A\times _{\cal G} \Gamma (W).
\nan

To get a concrete feel about $U$, we need to explain the geometry of
this bundle.  The metric on $Y$ and the hermitian metric
$\langle .\ , \ .\rangle$
on $\Gamma (W)$ naturally define a connection.
The Maurer-Cartan connection on
$\cal A \longrightarrow \cal A / \cal G $ is
flat while the hermitian connection on
has the curvature  $i \phi \mu \wedge \bar \mu $. This gives the expression
of term $i(\chi ,\mu ) \phi (\chi ,\mu)$ in (\ref{Thom}) in our case.

In our infinite dimensional setting, the map $C$ is given by
\ban
C: \quad \quad  & & \Omega ^0 (Y) \longrightarrow T_{(A, M)}(\cal A \times
\Gamma (W))\nonumber \\
& & \quad C(\eta )= (-d\eta, i \eta M),
\nan
and its dual is given by
\ban
C^* :\qquad & & \Omega^1(Y) \times \Gamma (W) \longrightarrow \Omega^0(Y), \\
& & C^*(\psi, N) = -d^* \psi +\langle N ,iM \rangle.
\nan
The one form $\langle\nu , \eta\rangle$ on $\cal A\times \Gamma (W)$ takes the
value
$$ \langle (\psi , N ), C \eta \rangle = \langle- d^* \psi ,\eta\rangle+
\langle N ,iM \rangle\eta$$
on the vector field $(\psi , N)$. We also easily obtain
$R (\lambda ) = -\Delta \lambda + \langle M, M\rangle\lambda$, where $\Delta =
d^*d $.
The $\langle\delta\nu , \lambda \rangle $ is a two form on $\cal A \times
\Gamma (W)$
whose value on $( \psi _1, N_1), (\psi_2 ,N_2)$ is $- \langle N_1 ,N_2
\rangle \lambda $.

Combining all the information together, we arrive at
the following  formula,
\ba
\pi _* \rho^* (U) &=& \int \exp \left\{ - \frac 12 |\rho |^2 + i (\chi,
\mu )\delta \rho + 2i \phi \mu \bar \mu\right. \nonumber \\
&+&\langle\Delta \phi, \lambda \rangle -\phi\lambda \langle M, M \rangle +i
 \langle N, N \rangle\lambda \nonumber\\
& &\left. +\langle \nu  ,  \eta \rangle \right\}
\cal D \chi\cal D \phi
\cal D \lambda \cal D \eta \cal b.  \label{Thom1}
\na
Note that
the 1-form  $i(\chi, \mu)\delta \rho $ on $\cal A \times \Gamma (W)
\times \Omega ^0(Y)$ contacted with the vector field $(\phi , N , b )$
leads to
\ba
& &2\chi^k\left[-\partial_k\tau + *(\nabla\psi)_k
-\Mb\sigma_k N-\Nb\sigma_k M\right] + \nonumber \\
 & & 2\langle \mu , \left[ i(\Dirac +b) N - (\sigma.\psi -\tau)M\right]\rangle;
\na
and the relation (\ref{relation}) gives  $|\rho |^2 = |* F -\bar M
\sigma M|^2 + |d b|^2 +|\Dirac M|^2 +b^2 |M|^2$.
Finally we obtain the Eular character
\ba
\pi_* \rho ^* (U)= \int \exp (- S) \cal D \chi\cal D \phi
\cal \lambda \cal D \eta
\cal{D}b   \label{Thom2}
\na
where $S$ is the action (\ref{S3}) of the three dimensional theory
defined on the manifold $Y$.

Integrating (\ref{Thom2}) over $\cal A\times_{\cal G}\Gamma(W)$
leads to Euler number
\ban
\sum_{[(A, M)]: s(A,M)=0} \epsilon^{(A, M)}.
\nan
which coincides with the partition function $Z$ of our
three-dimensional theory(recall that $Z$ is independent of
the coupling constant.).

\end{document}